\documentclass{PoS}

\usepackage{slashed}
\usepackage{epsfig, subfigure}

\def\ga{\mathrel{\raise.3ex\hbox{$>$\kern-.75em\lower1ex\hbox{$\sim$}}}}
\def\la{\mathrel{\raise.3ex\hbox{$<$\kern-.75em\lower1ex\hbox{$\sim$}}}}

\newcommand{\lam}{\lambda}
\def\lsim{\mathrel{\rlap{\lower4pt\hbox{\hskip1pt$\sim$}}
    \raise1pt\hbox{$<$}}}                
\def\gsim{\mathrel{\rlap{\lower4pt\hbox{\hskip1pt$\sim$}}
    \raise1pt\hbox{$>$}}}                

\title{Charged Higgs discovery potential in the single top mode in 2HDMs}

\ShortTitle{Charged Higgs in 2HDMs}

\author{Renato Guedes\\
        Centro de F\'\i sica Te\' orica e Computacional, Faculdade de Ci\^encias, Universidade de Lisboa, Av. Prof. Gama Pinto 2, 1649-003 Lisboa, Portugal. \\
        E-mail: \email{renato@cii.fc.ul.pt}}    
        
\author{Stefano Moretti \\
        NExT Institute and School of Physics and
Astronomy, University of Southampton Highfield, Southampton SO17
1BJ, UK.\\
        E-mail: \email{stefano@phys.soton.ac.uk}}       

\author{\speaker{Rui Santos}
\\
        Instituto Superior de Engenharia de Lisboa, Rua Conselheiro Em\'\i dio Navarro 1, 1959-007 Lisboa, Portugal \textit{and} \\
        Centro de F\'\i sica Te\' orica e Computacional, Faculdade de Ci\^encias, Universidade de Lisboa, Av. Prof. Gama Pinto 2, 1649-003 Lisboa, Portugal. \\
        E-mail: \email{rsantos@cii.fc.ul.pt}}

\abstract{We discuss the discovery potential of a charged Higgs boson in the single top mode. The models discussed include a CP-conserving
               and a CP-violating version of the softly broken $Z_2$ symmetric 2HDM potential. We conclude that the single top mode could help to
               constrain the ($m_{H^\pm}$, $\tan \beta$) plane in several versions of 2HDMs.}

\FullConference{Fourth International Workshop on Prospects for Charged Higgs Discovery at Colliders - CHARGED2012,\\
		October 8-11, 2012\\
		Uppsala Sweden}

\begin{document}

\section{Introduction}

\noindent
As the 8 TeV run at the CERN's Large Hadron Collider (LHC) is reaching the end, one may ask if a light charged Higgs will survive the confrontation with experimental data in extensions
of the Standard Model (SM) that allow for the existence of at least one charged Higgs boson. In models where portions of the parameter space will still survive the 8 TeV data, is there 
any hope of excluding a light charged Higgs for all of the parameter space by the end of the 13-14 TeV run?  
As will be clear later on, the answer to that question is no for some versions of two-Higgs doublet models (2HDMs). However,
we will show that a slight improvement can nevertheless be obtained by complementing the present search, based on the $t \bar t$ mode, with the search in the single top mode.

\section{Two-Higgs doublet models}

\noindent
CP-conserving as well as CP-violating (either explicit or spontaneous) 2HDMs' potentials with a softly broken $Z_2$ symmetry, $\Phi_1 \rightarrow \Phi_1$,
$\Phi_2 \rightarrow - \Phi_2$, can be written as
\begin{eqnarray}
V(\Phi_1,\Phi_2) &=& m^2_1 \Phi^{\dagger}_1\Phi_1+m^2_2
\Phi^{\dagger}_2\Phi_2 + (m^2_{12} \Phi^{\dagger}_1\Phi_2+{\rm
h.c}) +\frac{1}{2} \lam_1 (\Phi^{\dagger}_1\Phi_1)^2 +\frac{1}{2}
\lam_2 (\Phi^{\dagger}_2\Phi_2)^2\nonumber \\ &+& \lam_3
(\Phi^{\dagger}_1\Phi_1)(\Phi^{\dagger}_2\Phi_2) + \lam_4
(\Phi^{\dagger}_1\Phi_2)(\Phi^{\dagger}_2\Phi_1) + \frac{1}{2}
\lam_5[(\Phi^{\dagger}_1\Phi_2)^2+{\rm h.c.}] ~, \label{higgspot}
\end{eqnarray}
where $\Phi_i$, $i=1,2$ are complex SU(2) doublets.

Hermiticity of the potential forces all parameters except $m_{12}^2$ and $\lambda_5$ to be real.
The choice of $m_{12}^2$ and $\lambda_5$, together with the nature
of the vacuum expectation values (VEVs) will determine the CP nature
of the model (see~\cite{Branco:2011iw} for a review).  
This, in turn, dictates whether we end up with two CP-even Higgs states, 
usually denoted by $h$ and $H$, and one CP-odd state, usually denoted by $A$
(the CP-even case) or with three spinless states with undefined CP quantum number, usually denoted
by $h_1$,  $h_2$ and $h_3$ (the CP-violating case).
However, as long as the VEVs do not break the electric charge, which was shown
to be possible in any 2HDM~\cite{vacstab1}, 
there are in any case two (identical) charged Higgs boson states, one charged conjugated to the other. 

In this work we will focus on two specific realisations, one CP-con\-serv\-ing
and the other explicitly CP-violating~\cite{Ginzburg:2002wt,ElKaffas:2006nt}. 
In the CP-violating version
$m_{12}^2$ and $\lambda_5$ are complex and $Im (\lambda_5) = v_1 \, v_2 \, Im (m_{12}^2)$.
In both models the VEVs are real. By defining 
$\tan\beta=v_2/v_1$, it is then possible to choose the angle $\beta$
 as the rotation angle from the group eigenstates to the mass eigenstates in the 
 charged Higgs sector.
 By then extending the $Z_2$ symmetry to the Yukawa sector 
we end up with four independent 2HDMs, the well known~\cite{barger, KY} Type I (only 
$\phi_2$ couples to all fermions), Type II ($\phi_2$ couples to up-type quarks and $\phi_1$ couples to 
down-type quarks and leptons), Type Y or III ($\phi_2$ couples to up-type quarks and 
to leptons and $\phi_1$ couples to down-type quarks)  and Type X or IV ($\phi_2$ couples to all quarks and $\phi_1$ couples to leptons)
(details and couplings can be found in~\cite{Guedes:2012eu}).

We will now very briefly discuss the main experimental and theoretical constraints affecting the 2HDM parameter space.
The signal in our analysis originates from single top production with the subsequent decay $t \to b H^{\pm} \to b \tau \nu$.
Hence, only the charged Higgs Yukawa couplings are present and therefore 
the only parameters we need to be concerned with are $\tan \beta$ and the charged Higgs
mass.   
Values of $\tan \beta$ smaller than $O (1)$ together with 
a charged Higgs with a mass below 100 GeV are disallowed both by the constraints~\cite{BB} coming from $R_b$, from $B_q \bar{B_q}$ 
mixing and  from $B\to X_s \gamma$ for all models. Furthermore, data from $B\to X_s \gamma$ 
impose a lower limit of $m_{H^\pm} \ga 340$ GeV, but only for models Type II and Type Y.
The LEP experiments have set a lower
limit on the mass of the charged Higgs boson of 79.3 GeV  at 95\% C.L., assuming only 
$BR(H^+ \to \tau^+ \nu) + BR(H^+ \to c \bar s)=1$~\cite{LEP}.  
These bounds led us to take $m_{H^\pm} > 90$ GeV and $\tan \beta > 1$ for type I and X. We will also present results 
for model type II, where the bounds on the charged Higgs mass can be evaded due to the presence
of new particles as is the case of the MSSM.

\section{Results and discussion}

\noindent

$pp \to t \bar t$ is the best process to search for a charged Higgs 
boson at the LHC. However, because the single top production cross section
 is about one third of  $\sigma_{pp \to t \bar t}$, it 
certainly deserves a full investigation regarding its
contribution to the production of charged Higgs bosons. The signal consists 
mainly of a light charged Higgs boson produced via $t$-channel graphs,
$pp \to t  \,  j \to H^+ \, \bar  b \, j $ and $H^+  \to   \tau^+ \, \nu $,
together with 
$pp \to \bar t  \,  j \to H^- \,  b \, j $ and $H^-  \to   \tau^- \, \bar \nu $,
where $j$ represents a light-quark jet.
In what follows we are considering proton-proton collisions
at a center-of-mass (CM) energy of  $\sqrt{s} = 14$ TeV and a top-quark
mass $m_t = 173$ GeV.  We consider a charged Higgs boson mass interval of 90 to 160 GeV and the analysis
is performed in 10 GeV mass steps.

Maximising the signal-to-background significance
($S/\sqrt B$) makes both the $s$-channel and the $tW$ single-top production
modes negligible - only the $t$-channel process survives the set of cuts imposed.
Signal events were generated with POWHEG~\cite{Alioli:2010xd}
at NLO with the CTEQ6.6M~\cite{Nadolsky:2008zw} PDFs.
The top was then decayed in PYTHIA~\cite{Sjostrand:2006za}. We have considered
only the leptonic decays of the tau-leptons, that is, the signal final state is
$pp \to l \, b \, j \, \slashed{E}$, where $l=e, \mu$ (electrons and muons)
while $\slashed{E}$ means missing (transverse) energy.

The irreducible background, single-top production
with the subsequent decay $t \to b \, W^+$, was also generated with POWHEG.
The main contributions to the reducible background are:
 $t\bar t$ production, $W^\pm ~+$~jets (including not only light quarks and gluons,
but also $c$- and $b$-quarks) 
and the pure QCD background ($jjj$, where $j$ is any jet). 
The $t\bar t$ background was generated
with POWHEG while $W^\pm$ + jets (1, 2 and 3 jets) was
generated with  AlpGen~\cite{Mangano:2002ea}.
Finally, the QCD background was generated with 
CalcHEP~\cite{Pukhov:2004ca} (with CTEQ6ll PDFs). 
The hadronisation was performed with PYTHIA 6.  After hadronisation DELPHES~\cite{Ovyn:2009tx}, 
was used to simulate the detector effects.
For the detector and trigger configurations, we resorted to the ATLAS default definitions.

In order to maximise $S/\sqrt{B}$ we apply the following selection cuts (see~\cite{Guedes:2012eu} for details)

\begin{enumerate}
\item We demand one electron with $p_T > 30$ GeV or a muon with $p_T > 20$ GeV, and $|\eta| < 2.5$ for both leptons. 
\item We veto events with two or more leptons with $p_T > 10$ GeV. This
cut eliminates the leptonic $t \bar t$ background almost completely.

\item We veto events with leptons having $p_T$ above 55 GeV.

\item  Events with missing energy below 50 GeV are excluded. This is a cut that 
dramatically reduces the QCD background.

\item We ask for one and only one $b$-tagged jet with $p_T < 75$ GeV. 
We assume a $b$-tagging efficiency of  0.4 (with R = 0.7),
while the misidentification rates for the case of $c$-quark jets we take 0.1 and for lightquark/
gluon jets we adopt 0.01.

\item We reconstruct a "top quark invariant mass" as defined in~\cite{Guedes:2012eu} 
and demand all events to have this invariant mass above 280 GeV.

\item We define a leptonic transverse mass~\cite{Guedes:2012eu}, $M_T^{l \nu} $, and we have accepted events with 
$30 \, {\rm GeV} < M_T^{l \nu}  < 60 \, {\rm GeV} $ for charged Higgs masses between 90 and 130 GeV and 
$30 \, {\rm GeV} < M_T^{l \nu}  < 60 \, {\rm GeV} $ or $M_T^{l \nu} > 85 \, {\rm GeV}$ 
for higher values of the charged Higgs mass.

\item We have chosen events with one and one jet (non-$b$) only with $p_T>30$ GeV and $|\eta|\leq 4.9$.

\item We veto all events with a jet multiplicity equal to two or above for jets with $p_T>15$ GeV and $|\eta|\leq 4.9$.

\item We only  accept events where jets have a pseudorapidity $|\eta|\geq 2.5$.
\end{enumerate}

Putting all the numbers together we can find  $S/B$ and $S/\sqrt B$
as a function of the charged Higgs mass as presented in table~\ref{tab:signi}.
\begin{table}[h]
\begin{center}
\begin{tabular}{ccccc}
\hline
$m_H^{\pm}$ (GeV)   & Signal ($S$) & Background ($B$) & $S/B$ $(\%)$ & $S/\sqrt B$\\
\hline
90 & $38.6$ & $29.5$ & $130.92$ & $7.11$\\
100 & $40.5$ & $29.5$ & $137.19$ & $7.45$\\
110 & $45.6$ & $29.8$ & $153.00$ & $8.35$\\
120 & $47.7$ & $30.1$ & $158.26$ & $8.69$\\
130 & $42.3$ & $32.68$ & $129.53$ & $7.41$\\
140 & $117.1$ & $77.9$ & $150.25$ & $13.26$\\
150 & $120.0$ & $86.6$ & $138.64$ & $12.90$\\
160 & $109.7$ & $100.8$ & $108.81$ & $10.92$\\ 
\hline
\end{tabular}
\end{center}
\caption{Signal-to-Background ratio ($S/B$) 
and significance ($S/\sqrt B$) as a function of the charged Higgs mass for $\sqrt{s} = 14$ TeV and a luminosity of 1 fb$^{-1}$.
The 
numbers presented for the signal we take BR$ (t \to b H^{\pm}) = 100 \%$ and 
BR$ (H^- \to \tau^- \nu) = 100 \%$ and all other
BRs have the usual SM values.} 
\label{tab:signi}
\end{table}

\begin{figure}[here]
  \begin{center}
    \includegraphics[scale=0.37]{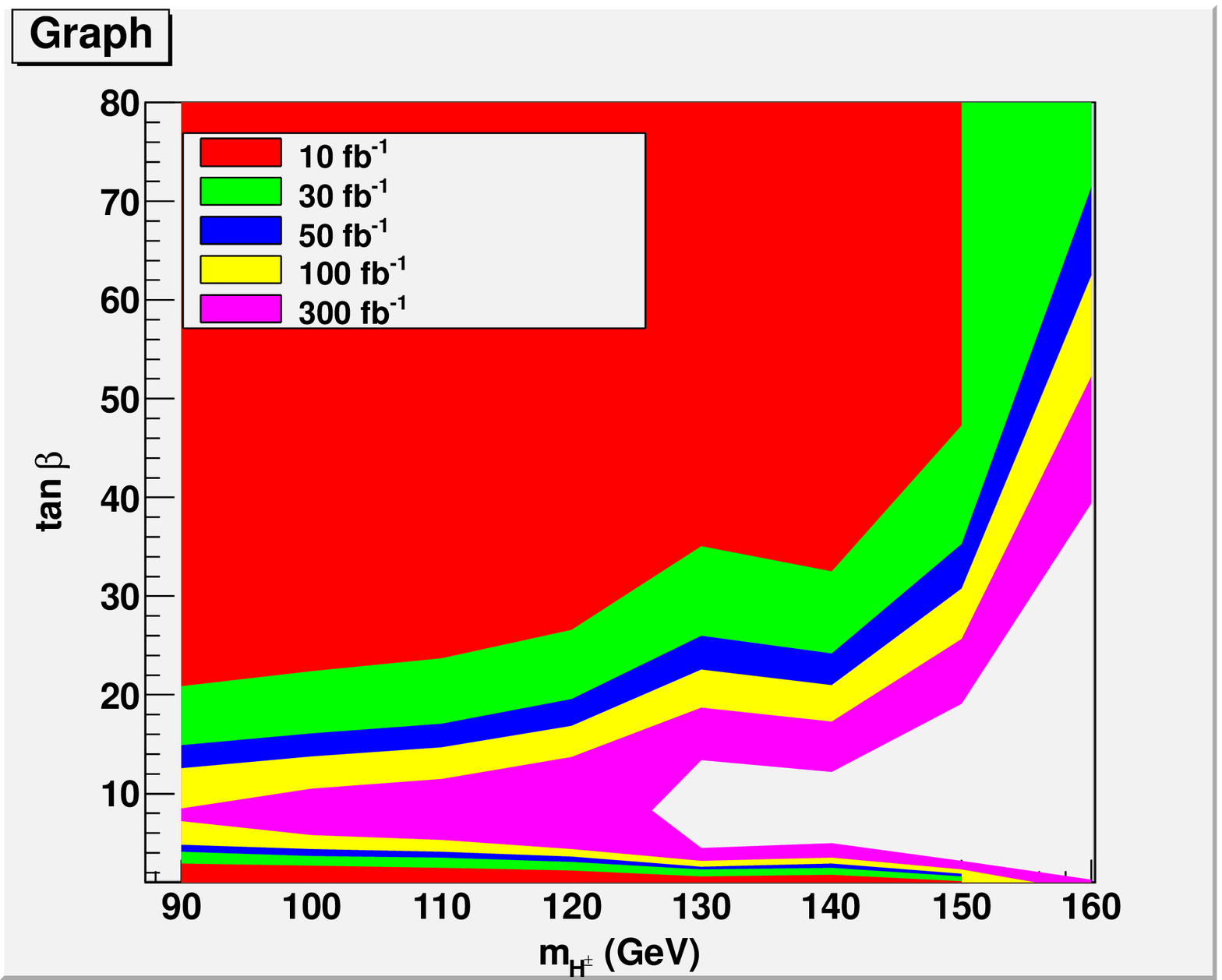}
    \includegraphics[scale=0.37]{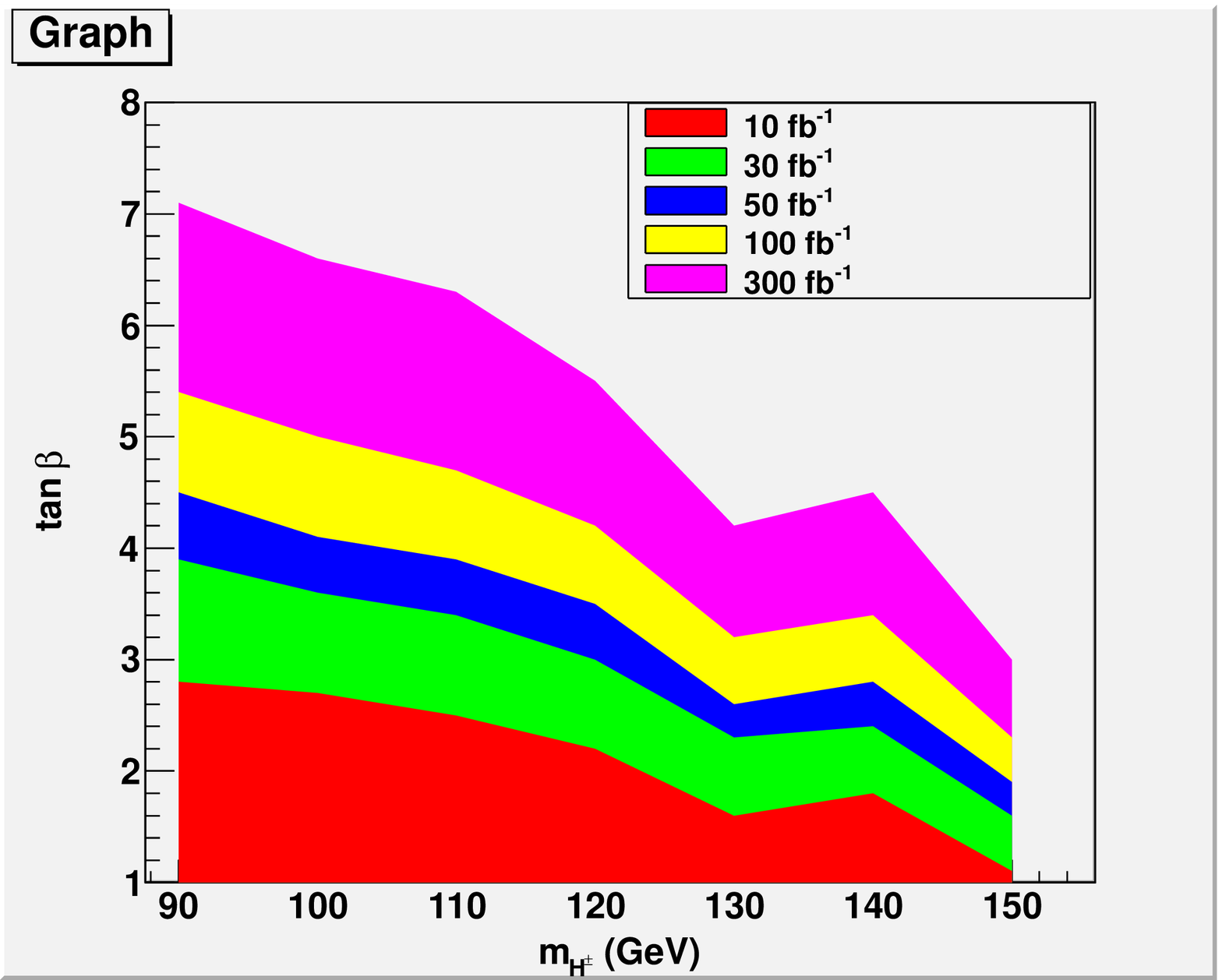}
    \caption{The MSSM (left) and Type X (right) exclusion limits over the ($\tan\beta$, $m_{H^\pm}$) plane at the 95\% CL
    assuming the LHC at 14 TeV and for several luminosity sets.}
    \label{fig10}
  \end{center}
\end{figure}

The results can be presented in a model independent manner~\cite{Guedes:2012eu}
and exclusion plots can be derived for the different 2HDMs.
In figure~\ref{fig10} we present the exclusion plots for the MSSM (left) and Type X (right) in 
the ($\tan\beta$, $m_{H^\pm}$) plane at the 95\% CL
assuming the LHC at 14 TeV and for several luminosity sets. 
%
%
%
%
%
The results show similar trends to the ones obtained for
$t \bar t$ production. We started with a cross section that
is about three times smaller than the $t \bar t$ one and ended up
with a result that is 2 to 3 times worse than the prediction
presented by ATLAS~\cite{Aad:2009wy}. 
It is expectable that both ATLAS and CMS
will improve the results on the single top mode presented here, improving
the constraints on the $(m_{H^\pm}, \tan \beta)$ plane.
One may now ask what are the chances to probe the entire $(m_{H^\pm}, \tan \beta)$ plane
by the end of the 14 TeV run. 
In view of the results for 7 TeV~\cite{ATLASICHEP, CMSICHEP}, one expects a type II charged Higgs to be excluded
by then. 
However, there are models
where the Yukawa couplings always decrease with $\tan \beta$ as is the case
of models I and X. For those models, we know that $pp \to t \bar t$ will provide
the strongest constraint on the $(m_{H^\pm}, \tan \beta)$ plane, and that the single
top mode is bound to contribute even if only with a slight improvement.
Are there any other processes that could help to probe the large $\tan \beta$ region?

There is another Yukawa process, $cs  \to H^{\pm}  (+jet)$~\cite{cs, Aoki:2011wd},  that could in principle help to probe the 
above mentioned region. It was however shown to be negligible for large $\tan \beta$.  
The remaining possibility~\cite{Aoki:2011wd} is to look for processes that
either do not depend on $\tan \beta$, or even better, that grow
with $\tan \beta$. There are terms both in $gg \to H^+W^-$ 
and in Vector Boson Fusion ($pp \to jj H^+H^-$ where $j$ is a light quark jet) that are independent of $\tan \beta$.
Furthermore, for the CP-conserving potential, there is a term in $gg \to H^+H^-$ that 
has the form 
\begin{equation}
\sigma_{pp \to H^+ H^-} \varpropto sin (2 \alpha) \, \tan \beta (m_H^2 -  M^2)
\end{equation}
where $\alpha$ is the rotation angle in the CP-even sector, $m_H$ is the heavier CP-even scalar mass
and $M^2 = m_{12}^2/(\sin \beta \, \cos \beta)$. Hence, there are regions of the 2HDM
parameter space that can be probed for larger values of $\tan \beta$. However, the bounds
will no longer be for a two parameter space but instead for a multi-dimension space 
with all 2HDM parameters playing a role. Further, values of the cross section
that could lead to meaningful significances, are only obtained for resonant production. Therefore,
only a small portion of the multidimensional space can be probed for large $\tan \beta$ (see~~\cite{Aoki:2011wd} for details). 

A final comment about theoretical bounds. Assuming that the Higgs boson was
discovered with a mass of 125 GeV, it was recently shown in~\cite{Maria}
that  for the particular case of an exact CP-conserving $Z_2$ symmetric model
$\tan \beta < 6$. Therefore, that particular model will probably see a light charged
Higgs ruled out when all the 8 TeV data is analysed.

\large{\textbf{Acknowledgments}}
SM is financed in part through the NExT Institute. 
The work of RG and RS is supported in part by the Portuguese
\textit{Funda\c{c}\~{a}o para a Ci\^{e}ncia e a Tecnologia} (FCT)
under contracts PTDC/FIS/117951/2010 and PEst-OE/FIS/UI0618/2011.
RG is also supported by a FCT Grant SFRH/BPD/47348/2008.
RS is also partially supported by an FP7 Reintegration Grant, number PERG08-GA-2010-277025.

\end{document}